\documentclass{mem}
\usepackage{natbib}\usepackage{txfonts}\usepackage{balance}
\usepackage{graphicx}
\usepackage[a4paper]{hyperref}
\begin{document}
\def\teff{$T\rm_{eff }$}
\def\kms{$\mathrm {km s}^{-1}$}

\title{
Distances to six Cepheids in the LMC cluster NGC1866 from the near-IR
surface-brightness method
}

   \subtitle{}

\author{
J. Storm\inst{1},
W. Gieren\inst{2},
P. Fouqu\'e\inst{3},
T.G. Barnes III\inst{4},
and M. G\'omez\inst{2}
          }

  \offprints{J. Storm}

\institute{
Astrophysikalisches Institut Potsdam --
An der Sternwarte 16, 14482 Potsdam, Germany
\email{jstorm@aip.de}
\and
Universidad de Concepci\'on, Concepci\'on, Chile
\email{wgieren@coma.cfm.udec.cl},
\email{matias@astro-udec.cl}
\and
Observatoire Midi-Pyr\'en\'ees, Toulouse, France;
\email{pfouque@ast.obs-mip.fr}
\and
The University of Texas at Austin, Texas, U.S.A.
\email{tgb@astro.as.utexas.edu}
}

\authorrunning{Storm et al.}

\titlerunning{Distances to Cepheids in NGC1866}

\abstract{
We derive individual distances to six Cepheids in the
young populous star cluster NGC1866 in the Large Magellanic Cloud
employing the near-IR surface brightness technique.  With six stars
available at the exact same distance we can directly measure the
intrinsic uncertainty of the method. We find a standard deviation
of 0.11 mag, two to three times larger than the error estimates
and more in line with the estimates from Bayesian statistical
analysis by Barnes et al. \cite{Barnes05}. Using all six distance estimates
we determine an unweighted mean cluster distance of $18.30\pm0.05$.
The observations indicate that NGC1866 is close to be at the
same distance as the main body of the LMC. If we use the stronger
dependence of the p-factor on the period as suggested by Gieren et
al. \cite{Gieren05} we find a distance of $18.50\pm0.05$ (internal error)
and the PL relations for Galactic and MC Cepheids are in very
good agreement.
}
\maketitle{}

\section{Observations} Storm et al. \cite{Storm05} have obtained radial
velocity curves as well as K-band light curves for  six Cepheids in the
young populous LMC cluster NGC1866. Combining these data with the optical
light curves from Gieren et al. \cite{Gieren00} we can apply the near-IR
surface brightness method as calibrated by Fouqu\'e and Gieren \cite{FG97}
to determine the distance to the individual stars.
\begin{figure}[t!]
\resizebox{\hsize}{!}{\includegraphics[clip=true]{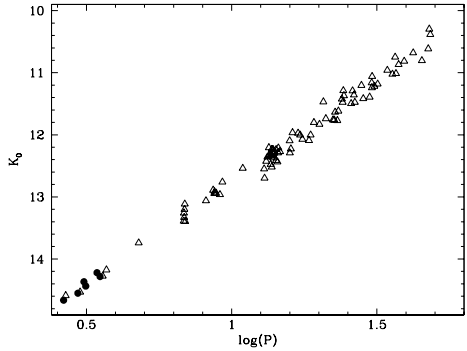}}
\caption{\footnotesize
Here we have over-plotted the dereddend mean $K$
magnitudes for the NGC1866 Cepheids (full circles) on
top of the mean $K$ magnitudes for the LMC field Cepheids
from Persson et al. \cite{Persson04}. The excellent agreement suggests
that NGC1866 is located close the LMC mean distance and 
that depth effects must be small if at all present.  This is in 
agreement with the model of van der Marel and Cioni \cite{vdmarel01} 
which suggests that the LMC disk at the location of NGC1866 
closer than the LMC barycentre by 0.06mag.
}
\label{logPK1866}
\end{figure}
\begin{table}
\caption{Individual distance moduli for the NGC1866 Cepheids using the
original and the new $p$-factor relations.}
\label{moduli}
\begin{center}
\begin{tabular}{lcccc}
\hline
\\
Star & $\log P$ & \multicolumn{2}{c}{$(m-M)$} & $\sigma$ \\
     &          &   org   & new  &    \\
\hline
\\
HV12199 &0.421 &18.34   &  18.55   &   0.09\\
HV12203 &0.470 &18.48   &  18.68   &   0.09\\
HV12202 &0.492 &18.29   &  18.49   &   0.07\\
HV12197 &0.497 &18.17   &  18.36   &   0.06\\
HV12204 &0.536 &18.20   &  18.39   &   0.04\\
HV12198 &0.547 &18.31   &  18.50   &   0.03\\
\multicolumn{2}{r}{Unweighted mean:} & 18.30  &   18.50  &   s.d. 0.11
\\
\hline
\end{tabular}
\end{center}
\end{table}

\begin{figure}[]
\resizebox{\hsize}{!}{\includegraphics[clip=true]{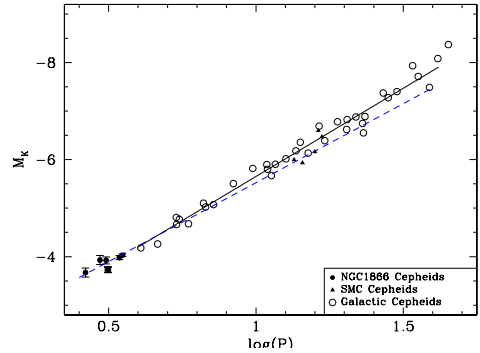}}
\caption{
\footnotesize
Here we have over-plotted the absolute magnitudes of
galactic Cepheids (open circles) on the NGC1866 Cepheids (filled
circles). The results are based on the Fouqu\'e and Gieren (1997)
calibration presented in Storm et al. \cite{Storm04} using a $p$-factor
which is only weakly dependent on period  [$p=1.39-0.03\log (P)$].
Note the excellent agreement between LMC and Milky Way stars. 
The dashed line represents the observed LMC relationship from 
Persson et al. \cite{Persson04} exactly as represented in
Fig.\ref{logPK1866} for an 
assumed distance modulus of 18.30 and illustrates the apparent
difference in slope between the LMC and galactic PL relations.
}
\label{logPMk}
\end{figure}

\section{Discussion}
Gieren et al. \cite{Gieren05} analyzed these stars together
with additional LMC Cepheids with longer periods using the
near-IR surface brightness technique. They found that the long
period stars also follow the observed relations in both of the
above diagrams, but not the dashed line in Fig.\ref{logPMk}. They also
found the unphysical result that the distances for these long
period stars were significantly longer than for the short period
NGC1866 stars. By introducing a stronger period dependence of the
$p$-factor which converts radial velocity into pulsations velocity
[$p=1.58 - 0.15\log (P)$] they could force the LMC stars to be at the
same distance independent of period.  As a consequence the slope
for the PL relation in Fig.\ref{logPMk} would become shallower leading to an
excellent agreement with the Persson et al. \cite{Persson04} relationship
(dashed line).  For the NGC1866 Cepheids this implies distance
estimates which are about 0.2 mag longer than with the original
$p$-factor relation.

\bibliographystyle{aa}

\end{document}